%% file: paper.tex
\documentclass[sn-apa]{sn-jnl}

\usepackage{graphicx}%
\usepackage{multirow}%
\usepackage{amsmath,amssymb,amsfonts}%
\usepackage{amsthm}%
\usepackage{mathrsfs}%
\usepackage[title]{appendix}%
\usepackage{xcolor}%
\usepackage{textcomp}%
\usepackage{manyfoot}%
\usepackage{booktabs}%
\usepackage{algorithm}%
\usepackage{algorithmicx}%
\usepackage{algpseudocode}%
\usepackage{listings}%
\usepackage{comment}

\theoremstyle{thmstyleone}%
\theoremstyle{thmstyletwo}%

\theoremstyle{thmstylethree}%

\begin{document}

%
\title{The Impact of International Collaborations with Highly Publishing Countries in Computer Science}


\author*[1]{\fnm{Alberto} \sur{Gómez-Espés}}\email{Alberto.Gomez-Espes@student.uibk.ac.at}

\author[2]{\fnm{Michael} \sur{Färber}}\email{michael.faerber@tu-dresden.de}
\equalcont{These authors contributed equally to this work.}

\author[3]{\fnm{Adam} \sur{Jatowt}}\email{adam.Jatowt@uibk.ac.at}
\equalcont{These authors contributed equally to this work.}

\affil*[1]{\orgdiv{Department of Information Systems, Production and Logistics Management}, \orgname{University of Innsbruck}, \orgaddress{\state{Tirol}, \country{Austria}}}

\affil[2]{\orgdiv{Institute of Computer Engineering}, \orgname{Dresden University of Technology}, \orgaddress{\state{Saxony}, \country{Germany}}}

\affil[3]{\orgdiv{Department of Computer Science}, \orgname{University of Innsbruck}, \orgaddress{\state{Tirol}, \country{Austria}}}


\abstract{This paper explores the dynamics of international collaborations in Computer Science, focusing on the roles of the three Publishing Superpowers, those are China, the European Union, and the United States. Leveraging a comprehensive literature review, we investigate collaboration patterns, research impact, retraction rates, and the impact of the Development Index on research outcomes. Our analysis reveals that while China, the EU, and the US play central roles in driving global research endeavors, other regions might shorten the distance between them in terms of the number of publications. Notably, collaborations involving these key regions exhibit lower retraction rates, underscoring their adherence to rigorous scientific practices. Furthermore, our findings highlight the influence of a country's Development Index on research impact, with \textit{Very High Developed} countries contributing to higher citation rates and lower retraction rates. Overall, our study sheds light on the significance of international collaborations in shaping the global research landscape and underscores the importance of fostering inclusive and ethical research practices for advancing knowledge and innovation in Computer Science.}




\maketitle

\section{Introduction}\label{sec1}


\textcolor{black}{International co-authorship has grown into a defining feature of modern scientific research, expanding significantly across countries and disciplines, particularly in science and technology fields since the 1990s \citep{leydesdorff2008international, wagner2017growth}. The appeal of international collaboration lies in its potential to amplify research impact by blending diverse perspectives and knowledge bases \citep{dusdal2021benefits, gomez2024benefits}. Studies indicate that such collaborations frequently yield higher citation rates and often enhance researchers’ access to resources and networks beyond what domestic partnerships might achieve \citep{puuska2014international, nomaler2013more}. Despite these benefits, international collaborations face challenges, including cultural and administrative barriers that can affect project efficiency and outcomes \citep{anderson2011international}.}

\textcolor{black}{This study focuses on Computer Science research, a field experiencing rapid global growth and expanding its influence across sectors, from academia to industry \citep{kostoff2001science}. Within this field, three primary regions—China, the European Union (EU), and the United States—play pivotal roles as global research leaders \citep{burke2022state}. Historically dominated by Anglo-American collaborations, the research landscape has become more inclusive, incorporating robust partnerships between Europe, North America, and the Asia-Pacific region \citep{gui2019globalization}. Although geopolitical tensions exist, China, the EU, and the US continue to collaborate extensively in engineering and technology fields, contributing substantially to Computer Science research \citep{jia2022impact, zhu2021analyzing}.}

\textcolor{black}{Alongside international collaboration patterns, socio-economic factors such as the Human Development Index (HDI) provide valuable context for evaluating the environments in which scientific work occurs. HDI, as a measure of a country’s overall development, captures key dimensions that support research infrastructure and education, impacting both the capacity for high-quality output and the formation of research networks \citep{undpreport}. Additionally, scientific integrity within this collaborative landscape is upheld through mechanisms such as retractions. Retractions serve as correctives in scientific publishing, yet differences in retraction rates across regions suggest that the socio-economic environment and local standards may play a role \citep{grieneisen2012comprehensive}.}

\textcolor{black}{In this study, we analyze the influence of these factors—international collaboration, HDI, and retraction rates—on Computer Science research. By examining publication patterns and research priorities across regions, we aim to gain insights into the dynamics of collaboration, the socio-economic influences on scientific output, and the disciplinary focuses that shape this rapidly evolving field.}

\textcolor{black}{In this research, we aimed to analyze whether regional differences are associated with variations in the number of published works, citations obtained, retraction rate, and research fields. Therefore, in this paper, we formulated the following research questions:}

\begin{itemize} \item \textcolor{black}{How are international collaborations in Computer Science related to publication retraction rates and scientific integrity, particularly in collaborations that involve or exclude China, the European Union, and the United States?}

\item \textcolor{black}{How does a country's Human Development Index relate to the quality and impact of scientific publications in Computer Science, and what differences are observed between Very High Developed and Low Developed countries?}

\item \textcolor{black}{Which topics have regions focused on, and how have research interests in Computer Science varied over the years across regions?} \end{itemize}

The findings from our research (see Section~\ref{discussion}) hold significant practical implications for the global scientific community. The dominant roles of the Publishing Superpowers in international collaborations underscore their pivotal contributions to the advancement of research, particularly within the field of Computer Science. While these regions play central roles in driving collaborative efforts and shaping research agendas, our analysis highlights the importance of diversifying partnerships to foster a more inclusive and robust research ecosystem. By engaging with a broader array of countries such as Canada, Australia, and Japan, researchers can leverage diverse perspectives and expertise, ultimately enriching the quality and relevance of research outputs on a global scale. For developing countries, by promoting inclusive and diverse research teams and fostering a culture of transparency and accountability, stakeholders can mitigate errors and enhance the reliability of scientific findings, ultimately advancing knowledge and innovation in Computer Science and attracting interest in collaborating with them. For China, the EU, and the US, findings can help them to make better decisions in terms of prioritizing collaborations with some regions over others.

The remainder of this paper is organized as follows: In Section~\ref{related_work} we discuss the related work on international scientific collaborations, as well as survey current literature about the position that China, the EU, and the US play in global research. In addition to this, we review current literature about the Human Development Index as a metric of socio-economic development and the actual findings about retraction in science. In Section~\ref{analysis}, we provide the details of our data source and explain the process we followed to collect and analyze the relevant information for our study. This section then shows the results we obtained when investigating the data. Finally, we discuss the results in Section~\ref{discussion}, draw conclusions based on the findings we got in Section~\ref{conclusion}, and outline potential future research directions in Section~\ref{limitations}.

\section{Related Work}\label{related_work}

\subsection{International collaborations}

International co-authorship is a world-extended trend that involves almost every country \citep{leydesdorff2013international}. It has been increasing since the 1990s in most STEM (science, technology, engineering, and mathematics) fields, and it continues its increase \citep{leydesdorff2008international, wagner2017growth, li2020research}. Many reasons embrace this kind of collaboration among researchers. Publications written collaborating internationally obtain higher citation rates on average than domestic collaborations \citep{puuska2014international, nomaler2013more, narin1991scientific}. They can expand researchers' knowledge, expertise, interests, and perspectives, and allow researchers to get funding for their research projects \citep{dusdal2021benefits, gomez2024benefits, anderson2011international}. International co-authorships usually also create resilient collaboration networks among researchers and can help young universities to obtain a higher impact in their publications \citep{franceschet2011collaboration, khor2016influence}. 

However, not everything is positive in international collaborations. The most evident and interdisciplinary barriers, such as miscommunication, management issues, language, cultural, and ethical differences can play a critical role in turning the collaboration into a failure \citep{anderson2011international, phelps2011international}. Additional factors may occur in the specific context of scientific research that can affect the result of the cooperation. In many cases, projects often lack clear goals, and it is challenging to define the expected outcomes and impacts of the research work. This makes it difficult to establish a clear set of indicators to measure success on all aspects \citep{boekholt2009drivers}. In addition to this, each country's specific context, such as graduate education level, language barriers or legal regulations can affect the result of the collaboration \citep{anderson2011international}.

Despite all the possible issues that research can face in the scope of international collaborations, they continue their increase, becoming a global trend,  important to consider by researchers and in the field of scientometrics in general.

\subsection{China, the European Union, and the US as research leaders in Computer science}

Although most countries participate in international collaboration in research, the US, some countries in Western Europe, and China were the most research-intensive regions \citep{leydesdorff2013international, burke2022state}. In the past, scientific collaborations used to be led by the Anglo-America sphere, but it has been changing gradually by a research world composed of Europe, North America, and Asia-Pacific. The US is usually seen as the top-level coordinator in the world, however, European countries such as the UK, Germany, Spain, or France as established science superpowers and play a relevant role in international work co-authorship \citep{gui2019globalization, jons2013global}. 

Even though current relationships between the three regions are not at their best moment, mainly between China and the US, collaborations between China, the EU, and the US in engineering and technology are commonplace and have strong linkages, with these three countries being the three largest contributors to each other \citep{jia2022impact, zhu2021analyzing, lee2020winners, wang2017network, wang2013international, yuan2018international}. Collaboration priorities among the regions are different, as well as the approaches taken to the goal of the partnership. For example, although the EU and the US universities tend to collaborate with the industry sector, Chinese researchers are less active in such kind of collaboration \citep{li2020status, gomez2024benefits}.

\subsection{Human Development Index}

\textcolor{black}{The Human Development Index (HDI) was introduced by Mahbub ul Haq in the Human Development Report of 1990, published by the United Nations \citep{undpreport}. The objective of the HDI was to redirect the attention of development economics from simply tracking national income to adopting people-centered policies. The intention was to establish a metric capable of evaluating a country's advancement not only based on economic growth but also on critical social outcomes. Currently, HDI is computed as the geometric mean of three indicators: life expectancy at birth, education (averaging mean years of schooling and expected years of schooling), and income (GNI per capita, PPP). The income component is presented on a natural logarithmic scale \citep{hickel2020sustainable}. The HDI has gained widespread recognition as a widely used indicator of human development, due to its reliance on observable indicators that can be meaningfully compared across different contexts. It is regularly promoted through the United Nations Development Program's annual reports. While other variables may better capture certain aspects of scientific collaboration, the HDI remains relevant due to its multidimensional approach to assessing development. Scientific collaboration often thrives in environments where not only economic resources but also social factors such as education and health are adequately supported. As such, HDI serves as a useful proxy for understanding the broader socio-economic environment in which scientific collaboration takes place. The education component of HDI, in particular, reflects a country's capacity to produce skilled researchers, while the life expectancy and income components provide insights into the overall well-being and resources available for long-term research efforts. The multidimensional nature of HDI helps capture the context in which research and collaboration are embedded, offering a more comprehensive view of the factors influencing scientific output. Although more targeted metrics (e.g., research funding, R\&D expenditure) could also provide a view of scientific infrastructure, HDI complements these by presenting a broader picture of human and societal development, which is critical for fostering sustainable collaboration networks. We acknowledge in the Limitations section that HDI does not fully capture all relevant factors, such as research funding, academic infrastructure, or specific indicators related to scientific output and collaborations. Additionally, several studies argue that HDI lacks important indicators such as ecological sustainability, political rights, and civil liberties \citep{sagar1998human, neumayer2001human, biggeri2018towards}. While HDI provides a valuable baseline, future research could benefit from incorporating additional variables that more specifically capture the nuances of scientific collaboration. Nevertheless, given its widespread use and recognition as a socio-economic development indicator, HDI offers a valuable framework for understanding how the broader developmental context might influence scientific cooperation and output \citep{ivanova1999assessment, biggeri2018towards}.}


\subsection{Retracted works}

Retraction serves as an intrinsic corrective mechanism within the scientific community, playing a crucial role in upholding and preserving the integrity of scientific literature \citep{chen2013visual}. Previous studies suggest that although most of the corrections made to articles were minor changes, the reasons for article retraction included ethical misconduct, scientific distortion, and administrative errors \citep{ajiferuke2020correction, al2019multiple, bar2018temporal}. In their study, \citet{ajiferuke2020correction} found that on average, the retracement process took approximately 587 days, and notably, certain retracted articles continued to receive citations even after the retraction occurred, which can lead to lower quality and credibility research \citep{rubbo2019citation, bar2018temporal, marco2021fraud}. In terms of self-citation, the retracted articles tend to obtain greater representativity than non-retracted articles \citep{bar2017post}. 

Retractions may vary depending on the fields. Medicine, Life Science, and Chemistry tend to obtain higher retraction rates than other fields such as Math, Physics, Engineering, and Social Sciences \citep{grieneisen2012comprehensive, sharma2021team, marco2021fraud, ribeiro2018retractions}. In the field of computer science, previous research suggests that retraction is a common outcome for publications that are related to ethical misconduct such as plagiarism, authorship issues, or publication duplication, and due to its vital importance for the research community, it should be taken more seriously \citep{shepperd2023analysis, al2019multiple, rubbo2019citation}. 

Apart from differences among the different research fields, retraction is also influenced by the participating countries in the collaboration, where developing countries tend to have more retractions than highly developed countries \citep{eldakar2023bibliometric}. Previous studies suggest that the internationalization degree of a country might impact the citation rates, where those that participated more in international collaborations tend to have a lower retraction rate \citep{zhang2020collaboration}. Countries such as the US, Germany, Japan, and the UK tend to have lower retraction rates compared to China, India, Iran, or South Korea \citep{he2013retraction, eldakar2023bibliometric}. In the case of the United States, although it is the country with the highest number of retractions, its retraction rate is three times lower than China's. However, in absolute terms, both countries have the highest number of retractions in the world \citep{zhang2020collaboration, ribeiro2018retractions}. Different factors can produce these differences among countries. The most common reasons that increase the retraction rate for some countries are the pressure for publication in an attempt to get better assessment in scientific evaluations, usually known by the expression "Publish or perish", relatively low costs of scientific integrity, and deliberate fraud \citep{lei2018lack, cavero2014computer}. 

The size of the collaborating group can also have an impact on the retraction rate, where small groups are more prone to retractions \citep{sharma2021team,siva2023retracted}. \citet{sharma2021team} set a threshold of ten participants and found that the majority of retractions are from teams composed of fewer individuals. Similar results were found by \citet{tang2020retraction}, who also found that the majority of the retractions are within the national borders. In our paper, we are considering these aspects but also extending them by providing geographical context, in particular in international research collaborations. Therefore, in contrast to them, we do not focus only on retractions, but on how retractions vary among the regions collaborating.

\section{Analysis}\label{analysis}

In this section, we first present how we collected and analyzed the data and then show the results obtained based on its analysis. We have divided our aspects into three different subparts: \textit{Number of Publications, Retracted Publications and Citations}, \textit{Impact in Human Development Index}, and \textit{Concepts}.

\subsection{Dataset}

Several options exist for accessing bibliometric information for this study from various data sources. Firstly, the Microsoft Academic Knowledge Graph (MAKG) is a substantial RDF dataset with over 210 million publications as of February 1, 2024, derived from the now-discontinued Microsoft Academic Graph (MAG). It encompasses a broad range of scientific publications and related entities, including metadata across multiple disciplines \citep{farber2019microsoft}.

Similarly, OpenAlex, a fully open scientific knowledge graph (SKG), was considered. It also originates from MAG and boasted 245 million works in its database as of February 1, 2024, expanding the pool of publications available for analysis \citep{priem2022openalex}. OpenAlex offers extensive metadata on publications, authors, and affiliations, which are critical for our analysis of research collaboration patterns and regional research priorities.

\textcolor{black}{Disciplinary coverage is a known variable in bibliometric data sources, as different databases emphasize certain fields over others. For instance, both MAKG and OpenAlex inherit disciplinary biases from MAG, where Computer Science, Engineering, and related technical fields are well-represented. To account for potential biases in disciplinary coverage, we reviewed studies on the disciplinary scope of OpenAlex and MAKG and assessed their suitability for our focus on Computer Science. OpenAlex, in particular, provides comprehensive Computer Science coverage due to its origins and the inclusion of extensive metadata for technical and applied research. This alignment with our area of study reduces potential coverage gaps that could impact our findings.}

In contrast, ArXiv was evaluated as another potential source. Although its data can be obtained in JSON format using datasets like Unarxive, it lacks detailed information about paper participants and their affiliations. Moreover, its relatively small corpus of just 2.4 million publications rendered it unsuitable for this study, leading to its discontinuation for our purposes \citep{Saier2023unarXive}.

We decided not to consider other data sources such as Web of Science, Scopus, or Google Scholar after reviewing the current state of the art, as these sources have already been extensively studied. Our focus on a different origin of data, such as OpenAlex, allowed us to both leverage unique filtering capabilities and explore under-researched data sources \citep{alryalat2019comparing, franceschet2010comparison}.

Ultimately, we chose OpenAlex as the primary data source for this study. Apart from the larger, up-to-date dataset compared to other alternatives, OpenAlex enables precise filtering, setting thresholds in searches, and thereby reducing processing time. The database’s detailed information about authors and institutional collaborations was also crucial for our analyses. Additionally, OpenAlex’s simplicity and easy integration with Python streamlined our data processing workflows, making it the optimal choice given our focus on disciplinary coverage within Computer Science.

\subsection{Data collection}

In the following section, we discuss the process we followed for the data collection. 

We obtained information on the works by utilizing OpenAlex API, filtering the works by “Computer science” concept and publication years from 1990 to 2021. OpenAlex uses an automated classifier trained on Microsoft Academic Graph corpus that tags each work with multiple concepts based on title, abstract, and the title of its host venue\footnote{https://docs.openalex.org/api-entities/concepts}. Apart from that, we only considered works that had authorship information filled, this means, that only those works whose participants were affiliated with an institution with a country location available were taken into account. There were some cases in which authors were affiliated to different institutions and in those cases we saved all the institutions linked to the author. We refer to these locations as "participant countries". Finally, we removed the publications in which there was only one author or all the authors were affiliated with institutions in the same country. We did that because the goal of this paper is to study international collaborations, therefore, national collaborations do not provide any value in this case. 

\subsection{Data preprocessing}

Once we had the raw data, we expanded it by adding the Human Development Index (HDI) for the year of publication of the work and attributed the value of the HDI to the same year. We gathered it from the Human Development Report\footnote{https://hdr.undp.org/data-center/human-development-index\#/indicies/HDI} provided by the United Nations. In our analysis, for simplicity, we considered Taiwan, Macao, and Hong Kong publications to be published by “China Mainland”, and the United Kingdom to be part of the European Union. Therefore, as we wanted to compare the results for the Publishing Superpowers and compare them to the rest of the world, we grouped our data based on the countries participating as follows:

\begin{itemize}
  \item Works, where only China, European Union countries, and the United States were collaborating between them, were tagged as "US-EU-CN". For such categorization, not all of them needed to participate in the collaboration, just 2 of them at least, and we did not consider collaborations between EU countries as international. 
  \item Publications where at least China, a country in the European Union or the US was collaborating with other countries were tagged as "Mixed".
  \item Works published as a result of a collaboration between authors not affiliated with an institution in China, any European Countries, or the US were tagged as "Other countries".
\end{itemize}

We refer to this classification as "Relation Group". It resulted in a dataset composed of 3,171,282 research papers.

\subsection{Publishing Superpowers impact on number of publications, retraction rates, and citations}\label{num_citations_results}

In this section, we aim to analyze how collaborations differ when Publishing Superpowers collaborate between them, with others and without them.

\input{table_number_of_publications_by_group}

\begin{figure}[tb]
    \centering
    \includegraphics[width=1\textwidth]{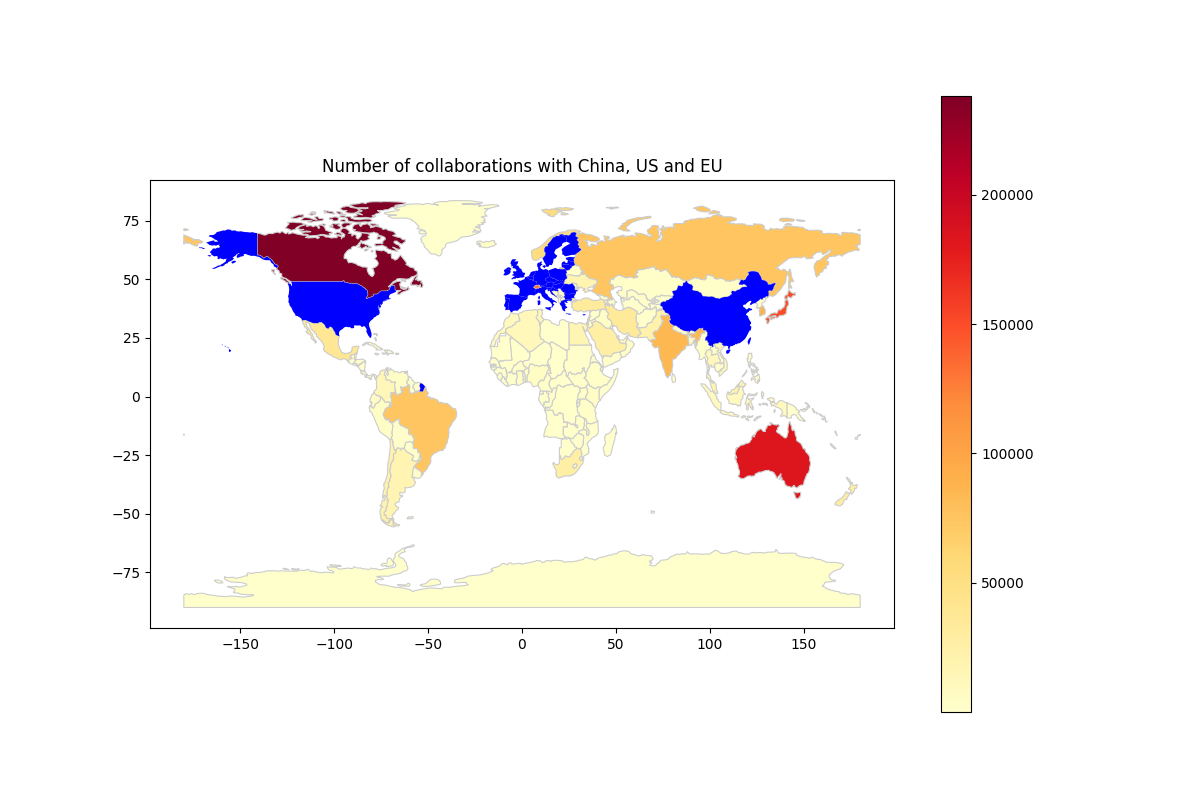}
    \caption{Number of publications in collaboration with the US-EU-CN}
    \label{fig:map_number_of_collaborations}
\end{figure}
\begin{figure}[tb]
    \centering
    \includegraphics[width=1\textwidth]{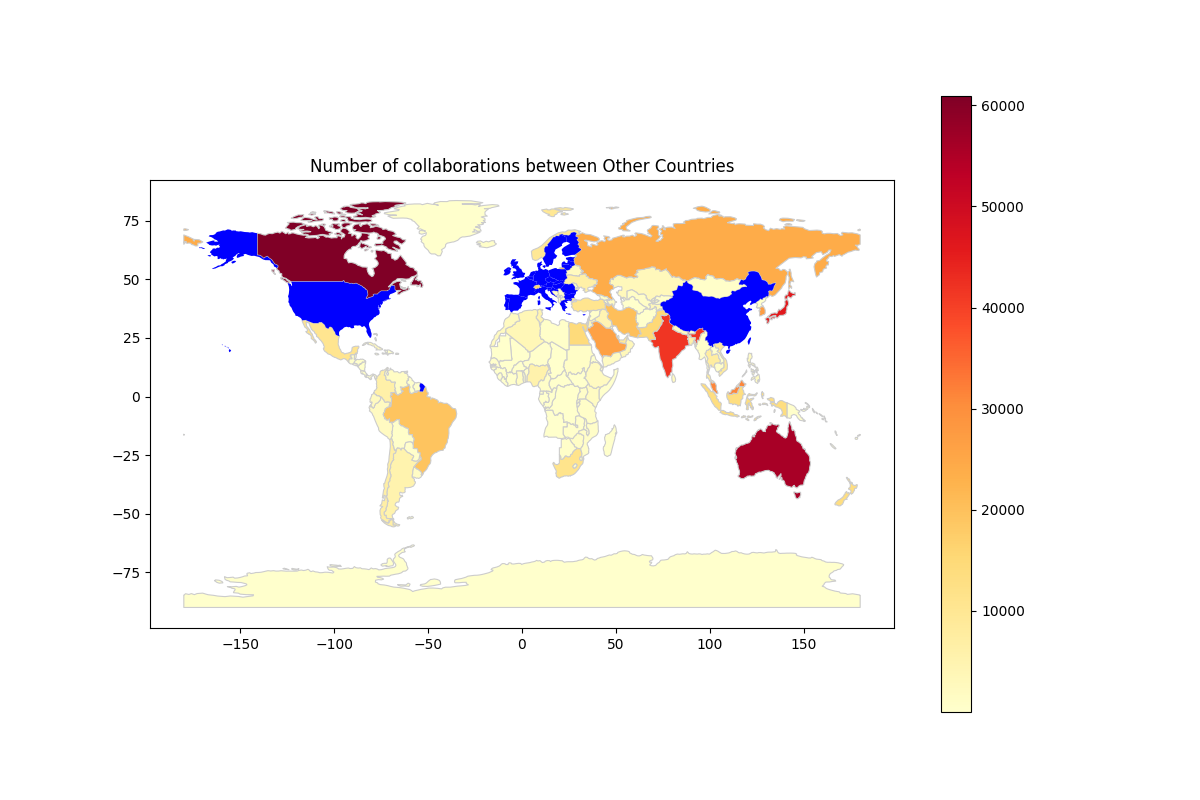}
    \caption{Number of publications in collaboration without the US-EU-CN}
    \label{fig:map_number_of_collaborations_other_countries}
\end{figure}

\subsubsection{Analysis on the number of coauthored publications}

First, we have analyzed the number of publications, citations, and retraction rates based on international clustering.  As we can see in Table 
{\ref{table:table_number_of_publications_by_group}} the group composed of either China, the EU, or the US, and another country was the one that published the most with 49.18\% of the publications. On the one hand, the second most prolific group was the one composed of China, the EU, and the US only with 40.98\% of the publications published just by collaborating between them. On the other hand, the rest of the countries in the world contributed 9.83\% of the total works done collaborating internationally. This result suggests that either China, countries from the EU, or the US participated in 90.15\% of the global international collaborations in the field of Computer Science. If we take a deeper look into Mixed countries collaborations we have discovered that the countries that collaborated the most with China, the EU, or China were Canada, Australia, Japan, Switzerland, and Korea, in order of the number of publications, from highest to lowest. Collaborations with these five countries represent 44.5\% of the total collaborations.

We obtained the countries that collaborated with China, the EU, and the US separately:
\begin{itemize}
\item China: Australia, Canada, Japan, Singapore, Korea, India, Pakistan, Saudi Arabia, Russia, Malaysia.
\item The European Union: Switzerland, Canada, Australia, Japan, Russia, Brazil, India, Norway, Israel, Mexico.
\item The United States: Canada, Korea, Japan, Australia, India, Israel, Switzerland, Brazil, Singapore, Russia.
\end{itemize}

In Figure \ref{fig:map_number_of_collaborations}, we can see a graphical representation of this cooperation, which Canada, Australia, and Japan highlight over the other countries. Then, we also got the collaboration data for the Other Countries group, and in Figure \ref{fig:map_number_of_collaborations_other_countries}, we can see that the 3 most collaborators were the same countries, Canada, Australia, and Japan, followed by India and Malaysia. These results suggest that Canada, Australia, and Japan are relevant partners for collaborating with, either for the US-EU-CN or other countries in the world.

\begin{figure}[tb]
    \centering
    \includegraphics[width=0.8\textwidth]{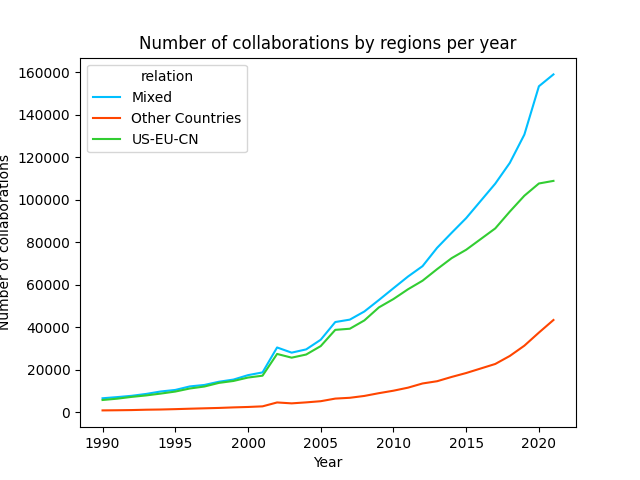}
    \caption{Amount of collaborations grouped Relation Group by Year}
    \label{fig:lineplot_collaborations_per_year_per_collaboration}
\end{figure}

If we review the results over the years, the number of collaborations between Other Countries has been the least productive if terms of the number of publications in all record data. However, the amount of collaborations for these groups seems to be quickly increasing since 2010. In Figure \ref{fig:lineplot_collaborations_per_year_per_collaboration}, we can see that the two most prominent publisher groups have been the Mixed group, followed by the one composed by China, the EU, or the US. The number of yearly partnered works seems to be similar between both groups from 1990 to 2002 when a decoupling process started until measured. This decoupling process was slow from 2002 to 2010, but it has been increasing since this year, in which the difference between the groups has been constantly increasing.
As a result, there was a difference of around 50,000 published works between the group composed of Mixed group, compared to the group composed of just those three regions collaborating. Results also suggest that the number of collaborations might have slowed down in the last years (2019-2021) in both cases. There seems to be a static tendency for collaborations within the US-EU-CN group, where the number of yearly collaborations neither increases nor decreases.

\subsubsection{Analysis on retraction rates}
 
We also considered the number of retracted papers for this study. Table {\ref{table:table_number_of_publications_by_group}} shows that in absolute values, the Mixed cluster was the one with the highest number of retracted works (648), followed by the US-EU-CN cluster (395) and the Other Countries (389). However, if we analyze the retracted works as a percentage of the total published works, we find that papers published by Other countries obtained a 0.12\% retraction rate. The Mixed international cluster received circa 3 times less retraction rate (0.04\%) and the works published by the US-EU-CN obtained a percentage of 0.03, having the lowest retraction rate of all the clusters.

The number of participants is another aspect of the collaboration that seems to affect the percentage of retractions. To measure it, we have obtained the percentage of retraction by the number of participants. Then, we performed a Shapiro-Wilk test to see if the data was normally distributed, obtaining that the data does not follow a normal distribution, $W$(172) = 0.296, $p-value$ = 0.000. As data was not normally distributed, we performed a Spearman test to check the correlation between the percentage of retractions and the number of participants. The result for the overall dataset suggested that there might be a moderate negative correlation between the variables, with a statistically significant correlation, $r$(170) = -0.543, $p-value$ = 0.000. Similar results were obtained when performing Spearman tests for the different groups independently. Collaborations without the US-EU-CN obtained the lowest correlation value, r(53) = -0.680, $p-value$ = 0.000, followed by collaboration between the US-EU-CN, $r$(106) = -0.592, $p-value$ = 0.000 and lastly, Mixed collaborations, r(170) = -0.494, $p-value$ = 0.000.  These results suggest that the retraction rate might decrease as the number of participants increases.

\subsubsection{Analysis on publications' citations}

If we look into the number of citations that the works received by the different groups, we find that publications done by Other countries (15.26) received more than 15 fewer citations on average compared to Mixed collaborations (25.82), and US-EU-CN collaborations (27.02) as shown in Table \ref{fig:table_number_of_publications_by_group}. We analyzed if there was a statistical difference between the groups by performing a One-way ANOVA method, suggesting that there is a statistical difference among the groups $F(2, 1559547) = 687.11$, $p = 0.000$. Furthermore, T-tests were conducted between the clusters, revealing statistically significant differences in all cases, with a $p-value$ $<$0.05. Finally, we can see in Table \ref{table:table_number_of_publications_by_group} that the median citations received by the US-EU-CN-only publications and the ones in which they participated with other countries are the same (6 median citations). On the other hand, publications, in which Other countries participated, received fewer citations when considering median citations (4 median citations).


\subsection{Results for the Human Development Index}\label{hdi_results}


Using the HDI scores provided for each country and year by the United Nations, we have analyzed how they impact in terms of collaborations. We have analyzed the group composed by the US-EU-CN by obtaining if there was at least a \textit{Very High Developed} country or at least a \textit{Low Developed} country. The thresholds for classifying the countries based on their HDI used are based on United Nations cutoffs \footnote{https://hdr.undp.org/reports-and-publications/2020-human-development-report/data-readers-guide}. Therefore, \textit{Very High Developed} countries were those whose HDI by the publication year was equal or above 0.8, and for classifying a country as \textit{Low Developed}, its HDI by the publication year had to be lower than 0.55. There were countries without information provided for the years studied, in those cases, the decision was to remove them and work with complete data available. The dataset analyzed is composed of 312,178 publications.

\begin{figure}[tb]
    \centering
    \includegraphics[width=1\textwidth]{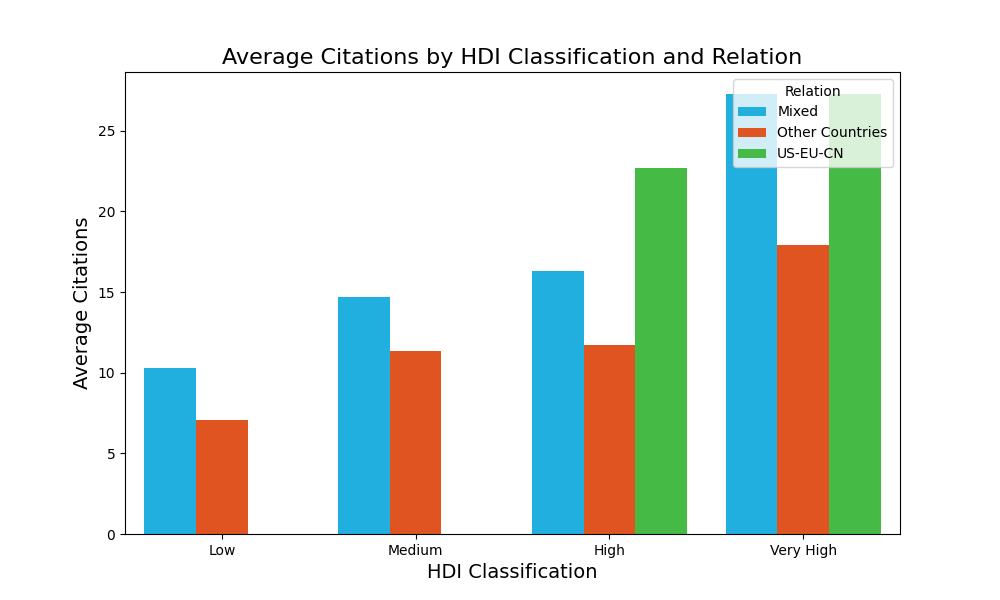}
    \caption{Average citations by average HDI and Relation Group}
    \label{fig:barplot_hdi_mean_citations_per_relation}
\end{figure}

In Figure \ref{fig:barplot_hdi_mean_citations_per_relation}, we see that in all the cases, the number of average citations decreases as the average HDI decreases. The results for the CN-EU-US cluster indicate that all countries are either \textit{High} or \textit{Very High Developed} countries, and in those clusters, their collaborations are the most relevant. For the Mixed and Other Countries cluster, we see that the differences in average citations between \textit{High Developed} and \textit{Medium} are not as huge as the difference between \textit{Very High} and \textit{High Developed}, and between \textit{Medium} and \textit{Low Developed}. This difference suggests that in terms of impact, the difference between collaborating with \textit{Medium} or \textit{High Developed} does not bring significantly different outcomes, compared to collaborating with \textit{Low} or \textit{Very High Developed}, which highly affects the number of citations received.

Once we obtained these results, we wanted to go deeper into \textit{Low} or \textit{Very High Developed} countries and see how they impact in their works.

\subsubsection{\textit{Very High Developed} countries collaborating}

We split the works into those that contained at least a \textit{Very High Developed} country collaborating in the paper, and those that did not. Works that had at least a \textit{Very High Developed} country participating received 26.03 citations on average. However, works that had no \textit{Very High Developed} country among their participants received 46\% fewer citations compared to works that had at least one \textit{Very High Developed} country. A t-test for analyzing the statistical difference was carried out, obtaining that the groups had statistically significant differences, $T$(1559548) = 38.21, $p-value$ = 0.000. We analyzed the amount of retracted publications in both groups. Works in which at least a \textit{Very High Developed} country participated had a retraction rate of 0.04\% (295 out of 1,532,059). On the other hand, the group without a \textit{Very High Developed} country had a retraction rate more than twice as high, having a 0.32\% retraction rate (87 out of 27,491).

\subsubsection{\textit{Low Developed} countries collaborating}

We split the works into those that contained at least a \textit{Low Developed} country collaborating in the paper, and those that did not. When analyzing works in which at least one participant country of the research was \textit{Low Developed} by the year of publication, we see that on average, works having a \textit{Low Developed} country received 19.18 citations, and the group without \textit{Low Developed} countries received 26.03 citations on average. After performing a T-test on this difference in the number of citations received, results suggest that there might be a significant difference between the groups, $T$(1559548) = -18.32, $p-value$ = 0.000. We have analyzed the retraction rates between the groups. Papers containing a \textit{Low Developed} country received a retraction rate of 0.14\% (64 out of 47328) and the papers without participant countries with Low HDI obtained a retraction rate of 0.04\% (584 out of 1,512,222), this corresponds to 36.8\% fewer retractions.


\subsection{Results for Computer Science Disciplines}\label{concept_results}

\begin{figure}[tb]
    \centering
    \includegraphics[width=0.7\textwidth]{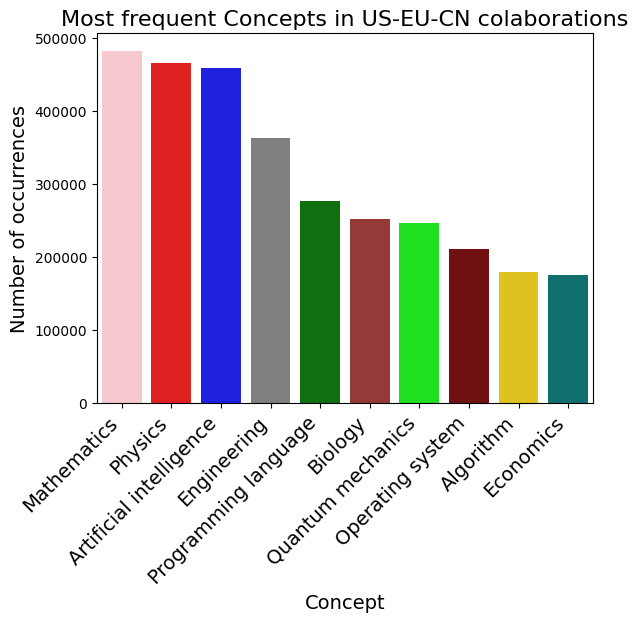}
    \caption{Most researched concepts by the US-EU-CN}
    \label{fig:cpt_in_us-eu-cn_colaborations_top_concepts}
\end{figure}

\begin{figure}[tb]
    \centering
    \includegraphics[width=0.7\textwidth]{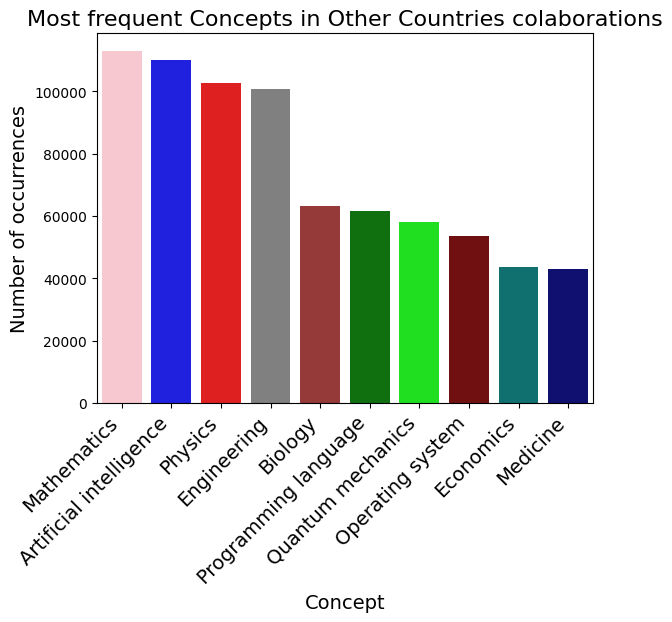}
    \caption{Most researched concepts by Other countries except US-EU-CN}
    \label{fig:cpt_in_other_countries_colaborations_top_concepts}
\end{figure}

\textcolor{black}{In this section, we analyze the concepts related to computer science publications as provided by OpenAlex. While our initial focus is on the field of computer science, it's important to note that OpenAlex uses an automated system to tag works with multiple concepts based on various data points like titles, abstracts, and venue names. This system sometimes assigns works to interdisciplinary fields due to overlapping areas of research and the hierarchical nature of concept tagging. Although our analysis primarily targets computer science publications, certain works may be tagged with concepts from related fields such as physics, biology, medicine, or economics. These disciplines often intersect with computer science, especially in areas like computational biology, data science, and interdisciplinary research involving AI and machine learning. For example, a paper on computational models in biology may be tagged under both biology and computer science. We filtered the works in advance by the "Computer Science" concept to ensure that all analyzed publications have a significant connection to the field.}
In collaborations within China, EU countries, and the US, the overall interests in overall have been Mathematics, Physics, Artificial Intelligence, Engineering, Programming languages, Biology, Quantum Mechanics, Operating systems, Algorithm, and Economics sorted by most frequent to less frequent concept as shown in Figure \ref{fig:cpt_in_us-eu-cn_colaborations_top_concepts}. For Mixed collaborations, Mathematics, Physics, Artificial Intelligence, Engineering, Biology, Programming language, Quantum Mechanics, Operating systems, Medicine, and Economics have been the most studied disciplines. Finally, as we can see in Figure \ref{fig:cpt_in_other_countries_colaborations_top_concepts}, for collaborations without the US-EU-CN collaborations, the most frequent concepts where Mathematics, Artificial Intelligence, Physics, Engineering, Biology, Programming language, Quantum Mechanics, Operating system, Economics, and Medicine. Results suggest that all the groups might have similar priorities when participating in international research collaborations, no matter the country they are collaborating with.


Furthermore, we wanted to know which topics have been relevant in last years. To achieve that, we obtained the 10 most common concepts in the last year for which we had data (2021), and we analyzed them over time.

\begin{figure}[tb]
    \centering
    \includegraphics[width=1\textwidth]{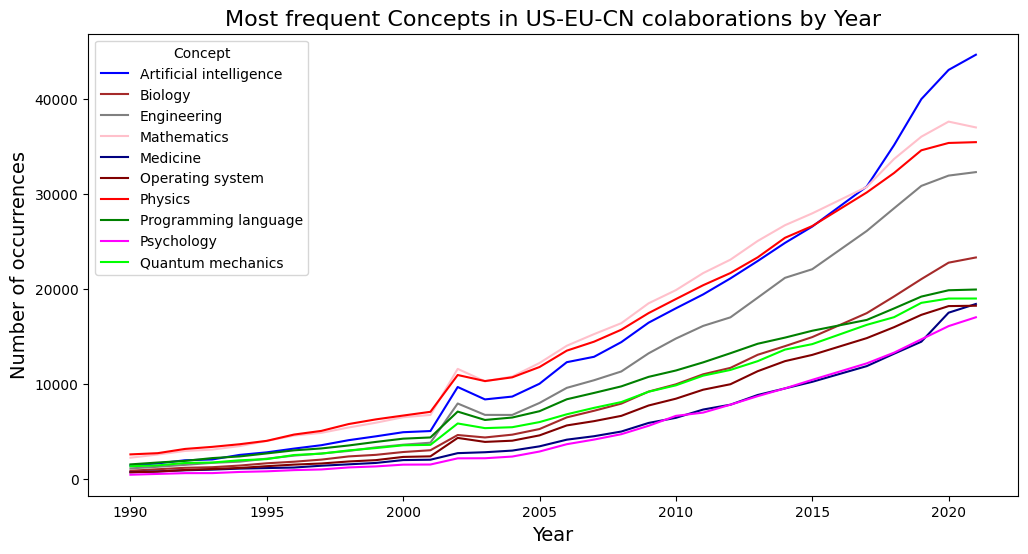}
    \caption{Current topics development during the time for US-EU-CN collaborations}
    \label{fig:cpt_in_us-eu-cn_colaborations_by_year_top_concepts_by_year}
\end{figure}

Figure \ref{fig:cpt_in_us-eu-cn_colaborations_by_year_top_concepts_by_year} shows that China, the EU, and the US have been constantly working on the concepts mentioned before. However, Mathematics and Physics have been surpassed by Artificial Intelligence since 2017, becoming the most researched topic. 
Other topics, such as Engineering, and Biology, have also obtained more interest over time. The first concept since the early 2000s, and Biology since the mid-2000s. In addition to this, we can see that Medicine received more interest since 2019. Another relevant aspect that we can notice is that some of the concepts have stopped increasing in the last 5 years. It means no more publications released than in the previous years, and in the case of Mathematics, Operating systems, and Quantum mechanics concepts, the amount of works published in collaboration between China, the EU, and the US has decreased. Apart from Artificial Intelligence, which is the leading interest, we can see 2 different groups of interests. The first one is composed of Mathematics, Physics, and Engineering, and the second one is composed of Biology, Operating systems, Quantum mechanics, Medicine, Psychology, and Programming language. 

\begin{figure}[tb]
    \centering
    \includegraphics[width=1\textwidth]{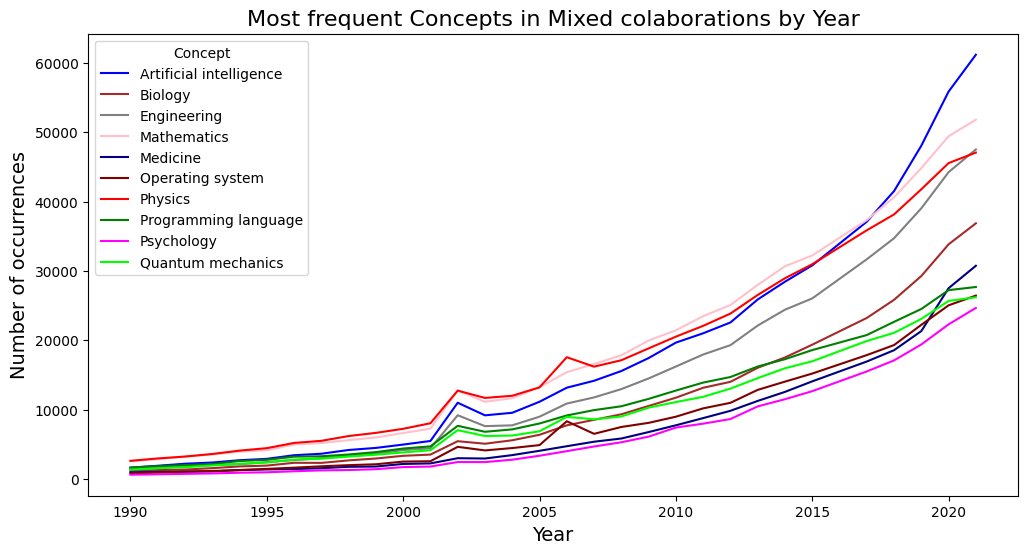}
    \caption{Current topics development during the time for Mixed collaborations}
    \label{fig:cpt_in_mixed_colaborations_by_year_top_concepts_by_year}
\end{figure}

In the case of collaborations between countries that do not involve China, the EU, or the US, we can see that in all cases there is a positive tendency from the 90s to 2021 in the number of collaborations focusing on topics shown in Figure \ref{fig:cpt_in_mixed_colaborations_by_year_top_concepts_by_year}. Similar to what happened to collaborations between China, the EU, and the US, Artificial Intelligence surpassed Mathematics in 2017. Although it is leading the research interest, the difference between it and the next most studied concepts is not that extensive compared to the collaborations between US-EU-CN. Therefore, we see 2 research interest groups, the first one composed of Artificial Intelligence, Engineering, Mathematics, and Physics. The second group is composed of Biology, Operating systems, Medicine, Machine Learning, Quantum mechanics, and Programming language. Similar to what happened in the previous case, Medicine increased its interest in 2019. 

\begin{figure}[tb]
    \centering
    \includegraphics[width=1\textwidth]{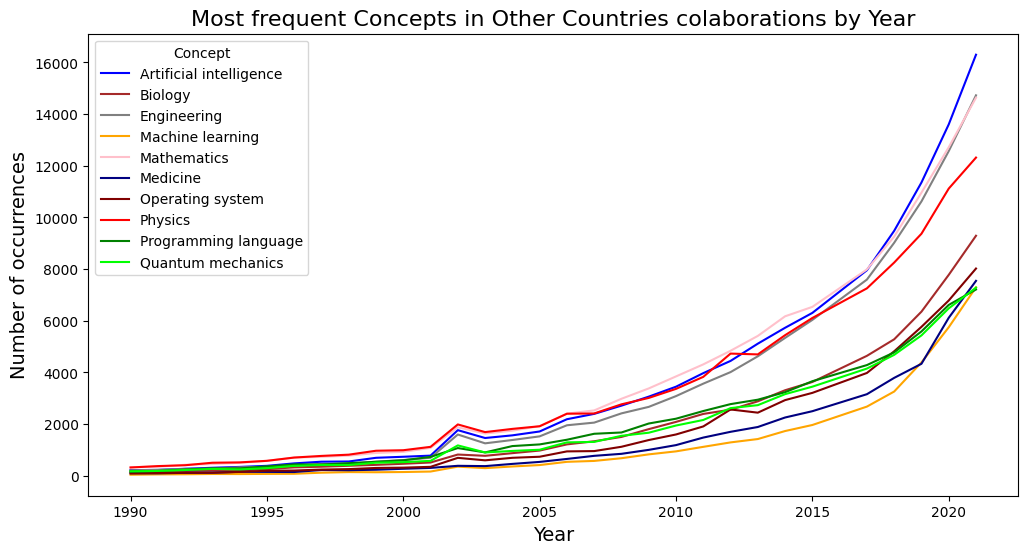}
    \caption{Current topics development during the time for collaborations without US-EU-CN}
    \label{fig:cpt_in_other_countries_colaborations_by_year_top_concepts_by_year}
\end{figure}

Finally, in Figure \ref{fig:cpt_in_mixed_colaborations_by_year_top_concepts_by_year}, we can see that the two previous tendencies obtained are combined when both groups collaborate. However, the results obtained by the group composed of China, the EU, and the US seem to be more aligned with the results obtained by the Mixed group. First, all the concepts are identical in both results, meaning that the most researched concepts in 2021 are aligned. Secondly, Artificial Intelligence is the most researched concept in 2021, also surpassing Mathematics and Physics, with 10,000 more occurrences than the second concept more researched. Thirdly, the concepts seem to be grouped more closely to the US-EU-CN pattern, although Biology is relatively more researched in the case of Mixed collaborations.

In Figure \ref{fig:cpt_in_us-eu-cn_colaborations_by_year_top_concepts_by_year} and Figure \ref{fig:cpt_in_mixed_colaborations_by_year_top_concepts_by_year}, we can see that the tendencies increased on an extraordinary basis for years 2002. After checking the data we found that the number of total works in the year 2002 increased by 31.11\% compared to 2001, which could explain the remarkable difference.

\section{Discussion}\label{discussion}

In this section, we explore the meaning and significance of the research results. 

First, we found that \emph{China, the countries from the European Union, and the US are playing a main role in international collaborations in the topic of Computer science}. Their participation rate in worldwide publications combining those that were done only collaborating between them and the ones that were done with other countries represent 90.16\% of total international published works. However, the difference between the yearly number of publications published in collaboration with the Publishing Superpowers, and those published by the Moderate Publishers could be reduced if the tendencies found in Figure \ref{fig:lineplot_collaborations_per_year_per_collaboration} keep going in the same direction as in the last years (2019-2021) as partially found by \citet{wagner2022changes} previously. Stagnation in the number of yearly publications partnered with researchers from other countries could lead to a loss of relevance for the US, the EU, and China, and a better position for the rest of the countries to fill the gap that these countries leave. 

If we take a look at relevant partners in co-authoring, we find that \emph{countries such as Canada, Australia, and Japan are relevant collaborators either with China, the EU, and the US or with other countries}. This position leaves these countries as relevant actors for doing research and collaborating, which might have a broad view of the Computer science field that could contribute by bringing different approaches to the different problems and issues that different parts of the planet face. In addition to this, small countries such as Switzerland or Singapore, seem to be relevant partners for the Publishing Superpowers, indicating that despite the differences in population or extension between the regions, those countries bring innovation and valuable knowledge.


\textcolor{black}{We found that publications by researchers not affiliated with China, the EU, or the US had three times higher retraction rates than those in which colleagues from these regions participated. This difference becomes fourfold if we only consider works solely from China, the EU, or the US. These findings suggest that researchers in these regions may play a significant role in preserving the integrity of scientific literature. The lower retraction rates observed when researchers from these three regions collaborate might indicate that they follow rigorous scientific practices. The reduction in retraction rates when they partner with other regions could imply that such good practices are shared, resulting in higher-quality and more trustworthy research. Moreover, the negative moderate correlation between retraction rates and the number of participants suggests that involving more researchers in a project might lead to more reliable work. This result aligns with previous findings by \citeauthor{sharma2021team}, and could be explained by the fact that a larger number of participants increases the likelihood of identifying gaps or flaws during the research process, thereby reducing the risk of retractions. However, it is important to interpret these results cautiously due to the relatively low number of retracted papers in the dataset, with only 1,432 retracted works out of 3,171,282 total works. The relevance of examining retractions in the context of international collaboration becomes particularly significant in fields like Computer Science, where global cooperation is crucial. Computer Science research often spans across institutions and countries, dealing with rapidly evolving technologies that require peer scrutiny from diverse perspectives to ensure rigor and accuracy. When researchers from China, the EU, or the US collaborate, they tend to bring well-established research practices and institutional frameworks that uphold high standards of data integrity, transparency, and reproducibility, which are critical in preventing scientific misconduct or errors that may lead to retractions. What makes the context of Computer Science particularly noteworthy is the pace of innovation and the broad applicability of its results. The field's focus on creating technologies that can be applied across diverse domains (e.g., machine learning in healthcare, cybersecurity for global systems) means that the impact of retracted research could have serious consequences, both scientifically and economically. Technologies developed in Computer Science often serve as foundational infrastructure for other scientific fields, as well as for industries ranging from finance to healthcare and logistics. Errors or unreliable findings in the core areas of Computer Science can thus ripple across many sectors, creating vulnerabilities in critical systems, disrupting technological progress, or even causing economic losses. Therefore, ensuring the reliability of publications through international collaborations helps reinforce the trustworthiness of the field.}


Higher retraction rate, combined with other factors such as less innovation or relevance in the field, 
might also explain the difference in citation rates on average between the groups. Publications without the participation of China, the EU, or the US received 41\% fewer citations on average than those in which they participated. This result suggests that those countries publish more relevant works than other countries. 

Moreover, the Human Development Index of the participant countries seems to be also relevant when measuring the impact (measured as the number of citations obtained by the publication)  that a scientific work generates. If there is a \textit{Very High Developed} country participating in the research, the publication received more citations on average and the retraction rate was lower compared to cases in which no \textit{Very High Developed} country participated. In addition to this, our results suggest that if a \textit{Low Developed} country participated in the research, publications received fewer citations, and the retraction rate was higher compared to works in which all the participants were at least medium-developed. These results suggest that the HDI, which includes indicators such as the expected years of schooling or the mean years of schooling, has a direct impact on the scientific outcomes that countries produce.

Finally, results regarding the concepts studied by the different groups suggest that Artificial Intelligence has been the main research topic from 2017 to 2021 for the three groups. Furthermore, all countries seem to be aligned in their research interests, having Mathematics, Physics, and Engineering as their most researched concepts, and only Psychology and Machine Learning the different concepts among the groups for the year 2021.  Curves between 2018 and 2021 for collaborations by China, the EU, and the US suggest that interest might have reached a maximum in capacity for collaborations between these three regions. This contrasts with the curves observed in collaborations in which those regions did not participate, which have a high positive inclination, suggesting that the interest in those concepts was increasing significantly. The increased interest in Medicine concept for the three groups starting in 2019 is also a notable aspect. This effect might be caused by the start of the COVID-19 epidemic, increasing the interest in this concept worldwide.

\section{Conclusion}\label{conclusion}

In conclusion, our comprehensive analysis of international collaborations in the field of Computer Science reveals several noteworthy trends and implications. The dominant roles played by China, the European Union, and the United States underscore their significant contributions to global research output. However, our findings caution against complacency, as the potential reduction in collaboration with other nations may diminish the influence of these leading regions. Stagnation in collaborative efforts could create an opportunity for other countries to emerge as key players in the international research landscape.

Our study reveals a notable difference in retraction rates among publications involving the Publishing Superpowers compared to those without. Collaborations with these regions show lower retraction rates, indicating their potential role in maintaining scientific integrity. Additionally, more participants are correlated with lower retraction rates, highlighting the importance of collaborative efforts in ensuring research reliability.

The impact of a country's Development Index on scientific publications is clear, with \textit{Very High Developed} countries contributing to higher citation rates and lower retraction rates. Conversely, involvement from \textit{Low Developed} countries correlates with lower citation rates and higher retraction rates, demonstrating the direct influence of Human Development Indicators on research quality and impact.

Our analysis of research focus areas reveals a shared emphasis on Artificial Intelligence across all groups from 2017 to 2021. However, differences in the most researched concepts for 2021 suggest evolving interests, with Medicine gaining prominence, possibly due to the global COVID-19 pandemic. The alignment of research interests in Mathematics, Physics, and Engineering highlights common ground among nations, fostering a collaborative and interconnected scientific community.

In summary, our study provides valuable insights into the dynamics of international collaborations in Computer Science, highlighting the roles of key nations, the importance of diverse partnerships, and the impact of development indicators on research outcomes. These findings contribute to a deeper understanding of the global scientific landscape, guiding future collaborative efforts and shaping the trajectory of research in Computer Science and related fields.

\section{Limitations}\label{limitations}

Data Source: The primary data utilized in this study stems from papers sourced from OpenAlex. While OpenAlex stands as a substantial academic database, it is important to acknowledge the possibility of divergent results arising from alternative data sources. As delineated in Section~\ref{analysis}, we did explore various data repositories; however, there remains the potential for disparities in findings when employing different databases or sources. 

Geographical and Cultural Considerations: The study opted to treat the European Union, China, and the United States as unified entities for the sake of simplification, potentially overlooking the intricate dynamics of international collaborations. Extensive literature highlights the substantial influence of barriers like geographical, cultural, and political distances on collaborative endeavors \citep{cerdeira2023international}. The diversity inherent in individual countries within this group can introduce distinct challenges to forging international research partnerships.

Limitations in usage of the Human Development Index in Scientometrics: The HDI primarily measures development indicators such as life expectancy, education, and income. While these factors are important, they might not fully capture the complexities of scientific collaborations, which can be influenced by various other socio-cultural, political, and economic factors. Particularly in science, the HDI may not capture specific indicators relevant to scientific collaborations, such as research funding, academic infrastructure, or scientific output. Thus, it might not provide a nuanced understanding of the scientific landscape.

Variability in Collaborative Dynamics across China, the European Union, and the United States: Collaborative endeavors within these three regions face distinctive challenges stemming from their inherent heterogeneity. The study did not thoroughly delve into these differences, which can significantly influence collaborative dynamics. Cultural, linguistic, and political variations within each region introduce complexities not typically encountered in more homogenous settings, shaping the nature of collaborations in unique ways. It is crucial to recognize these aspects when analyzing and interpreting the study's findings.

\section*{Declarations}

\textbf{Funding.} The authors did not receive support from any organization for the submitted work. \\
\textbf{Competing interests.} The authors have no competing interests to declare
that are relevant to the content of this article.

\begin{appendices}

\section{Collaborations with and without China, the EU and the US}

\input{mixed_collaborations_table.tex}

\end{appendices}

\bibliography{sn-bibliography}

\end{document}

%% file: table_number_of_publications_by_group.tex
\begin{table}[]
    \centering
    \begin{tabular}{lcccccccc}
    \hline
    International Cluster & Total works & Total retracted works & Average citations & Median citations \\  \hline
    US-EU-CN &  1,299,552 & 395 & 27.02 & 6.0\\
    Mixed & 1,559,550 & 648 & 25.82 & 6.0\\
    Other Countries & 312,180 & 389 & 15.26 & 4.0  \\
    \hline
    \end{tabular}
\caption{Published works, retracted works, average citations, and median citations grouped by international cluster.}
\label{table:table_number_of_publications_by_group}
\end{table}

%% file: mixed_collaborations_table.tex
\begin{table}[ht]
\centering
\begin{tabular}{lcccc}
\hline
Country & [1] & [2] & [3] & [4] \\
\hline
Canada & 238,240 & 60,926 & 35.52 & 20.23 \\
Australia & 182,731 & 55,763 & 34.70 & 21.97 \\
Japan & 151,262 & 45,944 & 23.72 & 13.57 \\
Switzerland & 130,752 & 16,715 & 37.03 & 22.86 \\
Korea, Republic of & 89,950 & 27,261 & 25.44 & 18.13 \\
India & 84,754 & 41,665 & 21.21 & 15.72 \\
Russian Federation & 75,333 & 24,275 & 20.54 & 7.86 \\
Brazil & 74,806 & 19,320 & 20.55 & 14.23 \\
Singapore & 64,718 & 15,827 & 32.92 & 24.15 \\
Israel & 59,370 & 8,334 & 36.35 & 22.16 \\
Norway & 53,537 & 9,702 & 28.89 & 18.35 \\
Mexico & 38,757 & 10,615 & 20.22 & 11.89 \\
Iran, Islamic Republic of & 34,984 & 20,335 & 22.58 & 18.76 \\
Turkey & 33,422 & 10,929 & 23.08 & 16.27 \\
New Zealand & 28,928 & 13,104 & 38.82 & 23.62 \\
Saudi Arabia & 26,538 & 26,260 & 30.35 & 15.61 \\
South Africa & 26,303 & 11,270 & 28.27 & 14.29 \\
Chile & 22,190 & 6,855 & 26.96 & 14.38 \\
Pakistan & 21,702 & 15,061 & 19.17 & 15.00 \\
Malaysia & 19,442 & 31,189 & 21.27 & 14.42 \\
Ukraine & 18,817 & 6,587 & 11.06 & 5.13 \\
Argentina & 17,829 & 5,378 & 24.62 & 13.83 \\
Egypt & 17,579 & 14,509 & 20.16 & 15.06 \\
Colombia & 13,788 & 6,240 & 19.96 & 9.67 \\
Tunisia & 13,153 & 4,802 & 11.82 & 9.90 \\
Algeria & 12,663 & 3,622 & 12.57 & 10.77 \\
Thailand & 12,357 & 7,766 & 21.99 & 11.68 \\
Viet Nam & 12,210 & 8,519 & 20.62 & 15.00 \\
United Arab Emirates & 11,335 & 9,123 & 20.56 & 14.82 \\
Serbia & 10,088 & 3,266 & 14.68 & 11.05 \\
Morocco & 9,703 & 2,222 & 13.97 & 10.77 \\
Indonesia & 9,454 & 13,443 & 15.11 & 7.22 \\
Qatar & 8,864 & 4,150 & 24.54 & 16.64 \\
Bangladesh & 7,115 & 6,662 & 17.58 & 13.92 \\
Lebanon & 6,561 & 1,902 & 23.35 & 13.10 \\
Nigeria & 5,604 & 5,558 & 17.31 & 10.17 \\
Ecuador & 5,227 & 2,328 & 14.82 & 5.71 \\
Jordan & 5,131 & 4,208 & 19.88 & 16.83 \\
Kenya & 5,037 & 1,727 & 31.65 & 14.20 \\
Philippines & 4,545 & 2,090 & 20.64 & 15.57 \\
Ghana & 4,493 & 1,510 & 19.17 & 11.41 \\
Iceland & 4,098 & 627 & 37.48 & 16.28 \\
Puerto Rico & 4,008 & 655 & 29.34 & 10.35 \\
Iraq & 3,961 & 4,570 & 15.43 & 12.86 \\
Venezuela, Bolivarian Republic of & 3,831 & 1,626 & 24.49 & 9.15 \\
Peru & 3,744 & 2,061 & 21.94 & 6.81 \\
Belarus & 3,561 & 2,344 & 13.08 & 5.49 \\
Cuba & 3,268 & 2,476 & 16.70 & 7.36 \\
Kazakhstan & 3,023 & 2,994 & 12.32 & 6.56 \\
Ethiopia & 2,980 & 2,009 & 24.35 & 11.86 \\
\hline
\end{tabular}
\caption{Highest 50 collaborators of CN-EU-US. Legend: [1] = Number of publications with CN-EU-US, [2] = Number of publications without CN-EU-US, [3] = Average citations with CN-EU-US, [4] = Average citations without CN-EU-US}
\label{table:mixed_collaborations}
\end{table}